\documentclass[%
superscriptaddress,
 amsmath,amssymb,
 aps,
floatfix,
prapplied, twocolumn
]{revtex4-2}
\usepackage{graphicx} 
\usepackage[margin=.84in]{geometry}
\usepackage{amsmath, amssymb}
\usepackage{tikz}
\usetikzlibrary{calc}
\usepackage{booktabs}
\usepackage{soul}
\usepackage[colorlinks=true]{hyperref}
\usepackage{multirow}

\newcommand{\normord}[1]{\vcentcolon\mathrel{#1}\vcentcolon}
\providecommand{\vcentcolon}{\mathrel{\mathop{:}}}
\newcommand{\ev}[1]{\ensuremath{\langle #1 \rangle}}

\newcommand{\mnoise}{\ensuremath{m_{\text{noise}}}}
\newcommand{\varnoise}{\ensuremath{s^2_{\text{noise}}}}
\DeclareMathOperator{\var}{Var}
\DeclareMathOperator{\cov}{Cov}
\DeclareMathOperator{\tr}{Tr}

\newcommand{\er}[1]{{\color{black} #1}}
\newcommand{\ernew}[1]{{\color{black} #1}}
\newcommand{\latest}[1]{{\color{black} #1}}

\newcommand{\err}[1]{{\color{black} #1}}

\begin{document}

\title{\latest{Witnessing quantum non-Gaussianity from intensity moments}}
\author{Éva Rácz}
\email{racz@optics.upol.cz}
\affiliation{Department of Optics, Palacký University, 17. listopadu 1192/12, 771 46 Olomouc, Czech Republic}
\author{László Ruppert}
\affiliation{Department of Optics, Palacký University, 17. listopadu 1192/12, 771 46 Olomouc, Czech Republic}
\author{Radim Filip}
\affiliation{Department of Optics, Palacký University, 17. listopadu 1192/12, 771 46 Olomouc, Czech Republic}

\begin{abstract}
   \latest{Direct measurement of quantum non-Gaussianity requires some variant of a discrete photon-resolving detection, which is feasible only for low mean photon numbers. For a large mean photon number, intensity detection by linear photodiodes provides a continuous signal; therefore, the Fock probabilities of the unknown input state are not directly available. On the other hand, intensity moments can be measured directly, and photon-number moments can be estimated. Therefore, we derive and analyze a quantum non-Gaussianity witness based solely on the photon number mean and variance (or alternatively, the second-order correlation $g^{(2)}$) of an unknown state. Due to the simplicity of the used photon-number moments, the measurement results are easy to correct for losses and additive noise. We provide examples of simple amplification-based measurement schemes where our witness can be applied directly, thereby opening pathways to proof-of-principle tests and applications.}
\end{abstract}

\maketitle

\section{Introduction}

Quantum non-Gaussian states, \ernew{that is, states that cannot be expressed as mixtures of Gaussian states,}  have a multitude of applications in metrology and quantum computations. 
In trapped ions and superconducting circuits, they are indispensable resources for realizing bosonic quantum sensors \cite{Wolf2019, McCormick2019, Wang2019, Pan2025, Rahman2025} over the Gaussian ones and bosonic quantum computing \cite{Lloyd1999, Gottesman2001, Michael2016} both for error correction  \cite{Ofek2016, Hu2019, Fluhmann2019, deNeeve2022, Sivak2023, Brock2025} and also because Gaussian states can be efficiently simulated on a traditional computer \cite{Mari2012}.


At many platforms, however, the preparation and detection of non-Gaussian states is still far from trivial, and therefore, simple and easily implementable criteria to verify quantum non-Gaussianity are very useful. 
This situation is especially limiting in quantum optics, as the light is control and probe of many other matter systems.
While certainly providing the necessary information, full state tomography does not qualify as either simple to perform or, in the absence of Wigner negativity, simple to evaluate. 

\ernew{Several witnesses of quantum non-Gaussianity have been proposed in the literature, tailored to specific practical scenarios. Generally, they rely on identifying observable statistics that are unattainable by any mixture of Gaussian states. For example, Ref.~\cite{Genoni2013} defines a witness based on the value of the Wigner function at the origin of the phase space \(W(0,0)\) and the mean photon number \(m\) and proves that \(W(0, 0)\) cannot be lower than a certain function of \(m\) for any mixture of Gaussian states. Similarly, the single-photon probability \(p_1\) cannot be lower than a function of the vacuum probability \(p_0\) for any mixtures of Gaussians \cite{Filip2011, Jezek2011}. The latter approach has been thoroughly studied and extended over the years \cite{Jezek2012_Experimental, Straka2014, Straka2018}, \ernew{including a witness based on the mean photon number beside the vacuum probability \(p_0\) \cite{Fiurasek_2021}}. Other works rely on alternative quasiprobability distributions \cite{Hughes2014_QNG_in_phase_space, Kuhn2018_QNG_and_quantification_of_NC}, generalized squeezing \cite{Brauer2025_GeneralizedSqueezing}, expectation of the operator \(e^{-c\hat x^2}+e^{-c\hat p^2}\) as a function of \(c\) \cite{Happ2018}. \er{Most recently, a procedure for detecting non-Gaussianity in cluster states based on non-Gaussian nullifiers has been proposed in \cite{Kala2025}.}

The notion of quantum non-Gaussianity has been extended to form hierarchies (with the corresponding witnesses) \cite{Lachman2019_NGhierarchy, Chabaud2021, Fiurasek2025}: a state possessing \(n\)-photon non-Gaussianity is a state that cannot be expressed as a mixture of squeezed displaced \er{lower-order} Fock states. In this sense, the notion of quantum non-Gaussianity discussed in this work corresponds to single-photon non-Gaussianity.

}





In experimental schemes involving large-gain amplification to compensate for subsequent loss and noise, individual Fock probabilities might be inaccessible due to the magnitudes of intensities created. 
However, photon number moments \er{after and before} amplification can be estimated (including loss- and noise correction) from the integrated-intensity moments \err{\cite{Perina1968, Perina2010, Perina2017}}, which provides a strong motivation to derive a non-Gaussianity witness based on moments only.
Simultaneously, a single-mode pump of the amplifier effectively filters other modes of the signal, therefore, the output state is close to perfect single mode.    
Regarding non-classicality, multiple witnesses have been formulated in terms of photon number and integrated intensity moments \cite{Perina2017}; however, similar criteria
for non-Gaussianity are missing from the literature. In this manuscript, therefore, we ab-initio introduce a quantum non-Gaussianity (QNG) witness in terms of the photon number mean and variance. 
Analogously to the probability-based witness, for a fixed value of the mean photon number \(m \equiv \langle \hat N \rangle\), the photon number variance \ernew{\(s^2 \equiv \langle (\hat N - m)^2\rangle \)} cannot go below a certain value (dependent on \(m\)) if the state is a single-mode mixture of Gaussian states. Due to the fact that it is based on the first two moments, the new witness behaves favorably regarding additive noise and loss, which is easy to exploit.

First, we provide the analytic form of this new moment-based witness and provide proof of its validity. Then, we suggest a few experimental schemes involving amplification where the new witness is convenient to use. Finally, we \ernew{analyze the applicability of the new witness for lossy Fock states and lossy photon-added thermal states, and discuss how to treat loss and additive noise in general}.


\section{Moment-based QNG witness}\label{sec:results}

Mixtures of Gaussian states in a single mode can be represented through their Wigner functions as
\[W(x,p) = \int \mathrm dP(\lambda)\, W_\lambda(x, p),\]
for some index variable \(\lambda\) and \(P(\lambda)\) denoting a probability measure on the set of possible values of \(\lambda\). The functions \(W_\lambda(x, p)\) are pure Gaussian states obeying the Heisenberg uncertainty principle: 
\[W_\lambda(x, p)\propto \exp\left\{-\frac 12
\left(\begin{array}{c}
x-x_\lambda\\
p-p_\lambda
\end{array}\right)^{\mathrm T}
C_\lambda^{-1}
\left(\begin{array}{c}
x-x_\lambda\\
p-p_\lambda
\end{array}\right)
\right\},\]
where \(x_\lambda\) and \(p_\lambda\) denote mean quadratures and \(C_\lambda\) the covariance matrix of the two quadratures, 
\(\sqrt{\det C_\lambda} \geqslant 1/4\).

For such states, the first two photon number moments can be calculated as
\begin{align}
\label{eq:Wigner-mean}\ev{\hat N} &= \int \mathrm dx \mathrm dp\, \left(x^2 + p^2-\frac 12\right)W(x, p)  \\
\nonumber &= \int \mathrm dP(\lambda)\, \int \mathrm dx \mathrm dp\, \left(x^2 + p^2-\frac12\right)W_{\lambda}(x, p)\\ 
\nonumber  &=\int \mathrm dP(\lambda)\, \ev{\hat N_\lambda},\\
\label{eq:Wigner-var}\ev{\hat N^2} &= \er{\int \mathrm dx \mathrm dp\, \left[(x^2 + p^2)^2-(x^2+p^2)\right]\, W(x, p)}\\
\nonumber &= \int \mathrm dP(\lambda) \int \mathrm dx \mathrm dp
(x^2 + p^2)^2W_{\lambda}(x, p) \\
\nonumber &- \int \mathrm dP(\lambda)\, \ev{\hat N_\lambda} - \frac 12\\
\nonumber &=\int \mathrm dP(\lambda)\, \ev{\hat N_\lambda^2},
\end{align}
where \er{for the first equality of \eqref{eq:Wigner-var},} we used the identity that the symmetrically ordered second moment can be given as \((\hat a^{\dagger 2} \hat a^2)_S = \hat N^2+\hat N+1/2\).
That is, for mixtures, non-centered photon number moments are the appropriate averages of the component moments. 
Note that the second moment defined in \eqref{eq:Wigner-var} does not involve ordering, that is, \(\ev{\hat N^2} = \ev{\hat a^\dagger \hat a a^\dagger \hat a}\).
We will use the notations \(m \equiv \ev{\hat N}\) and \(s^2 \equiv \ev{\hat N^2} - m^2\) for the photon number mean and variance, respectively. Note that although the photon number variance cannot be, in general, calculated as the average of the component variances, given a fixed value of \(m\), minimizing \(s^2\) is equivalent to minimizing the non-centered second moment \(\ev{\hat N ^2}\). 

For dim states (states for which the probability of observing three or more photons is negligible), the Hanbury Brown--Twiss coincidence scheme provides a straightforward way to estimate vacuum and single photon probabilities \(p_0\) and \(p_1\). For any given value of \(p_0\), there is a maximal value of \(p_1\) that a mixture of Gaussian states can possibly attain \cite{Jezek2011}. Therefore, a measurement result where the single-photon probability exceeds this limit indicates quantum non-Gaussianity.

Similarly, we will show that mixtures of Gaussian states in a single mode have a minimal photon number variance for a fixed mean photon number. This minimum-variance curve can be given by the following parametric equation:
\begin{align}
\label{eq:m}m_{\mathrm{NG}}(r) &= \frac{e^{6r}}{4}-\frac 12 + \frac{e^{-2r}}{4}\\
\label{eq:s2}s^2_{\mathrm{NG}}(r) &= \frac{3e^{4r}}{8}-\frac 12 + \frac{e^{-4r}}{8},
\end{align}
with \(r \geqslant 0\). 
In other words, there are pairs of values of \(m\) and \(s^2\) inaccessible to mixtures of Gaussian states \ernew{(area below the orange line in Fig.~\ref{fig:squeezed-coherent})}, and this can be used to witness the non-Gaussianity of quantum states of light even if their Wigner function is strictly positive. While a bit cumbersome, we can also express the analytic witness of quantum non-Gaussianity in a non-parametric form as
\begin{align}
\nonumber m  &>  \frac{\left(4 s^2+\sqrt{16 s^2 (s^2+1)+1}+2\right)^{3/2}}{12 \sqrt{3}}\\
\label{eq:QNG-nonparam}&+\frac{\sqrt{3}}{4\sqrt{4 s^2+\sqrt{16 s^2 (s^2+1)+1}+2}}-\frac 12.
\end{align}

\ernew{As detailed in the introduction, many witnesses of quantum non-Gaussianity have been developed, each applicable in different scenarios. Since neither of these is based solely on photon number moments, and also because the use cases are different, providing a fair comparison to the proposed witness is not straightforward. 
Nevertheless, for the sake of simplicity, we transformed the vacuum- and single-photon-probability-based witness from \cite{Jezek2011} to photon number mean \(m\) and variance \(s^2\) (see Appendix \ref{sec:conversion} for derivation).
\begin{figure}[!t]
\centering
\begin{tikzpicture}
\node[anchor=north east, inner sep=0] (top-left) at (0,0){\includegraphics[width=\linewidth]{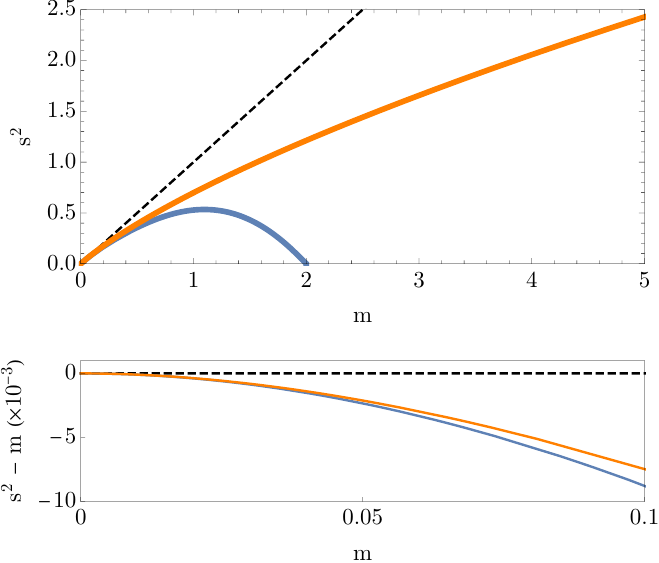}};
\node[inner ysep=10, inner xsep = 35, anchor=north west] (a) at (top-left.north west){\footnotesize(a)};
\node[inner ysep=28, inner xsep = 35, anchor=south west] (b) at (top-left.south west){\footnotesize (b)};
\end{tikzpicture}
\caption{Comparison of QNG witnesses. (a) In terms of photon number mean and variance. The blue line shows the border of the converted Fock probability-based witness, and the orange line the new moment-based witness: the latter has an extended applicability towards bright states. The areas below either line correspond to pairs of values indicating quantum non-Gaussianity. The dashed black line shows the border of the non-classicality witness, \(s^2<m\). 
(b) The same lines in terms of \ernew{\(s^2-m\) and \(m\)}
for a narrower range of mean values, showing that the two witnesses are equivalent for low values of \(m\). \label{fig:squeezed-coherent}}
\end{figure}
Figure \ref{fig:squeezed-coherent} shows this transformed witness with a blue line.
{
From Fig.~\ref{fig:squeezed-coherent}(b) we can see that the two witnesses become} equivalent as \er{\(m\ll 1\)}: using the right-hand side of \eqref{eq:QNG-nonparam}, we obtain that
\[m_{\mathrm{NG}} = s^2_{\mathrm{NG}}+s^4_{\mathrm{NG}}+\mathcal O (s_{\mathrm{NG}}^6),\]
which is the same as the low-mean behavior of the probability-based witness \cite{Jezek2011} if \(p_{3+} = 0\) is assumed. 
\er{However, for larger values of the mean photon number \(m\), the moment-based witness considerably relaxes the condition for identifying QNG. Note that this is not necessarily a fair comparison (the transformation of the probability-based witness is only exact if \(p_{3+} = 0\)); you will find a better comparison in terms of non-Gaussianity depth in Sec.~\ref{sec:resilience_to_loss_and_noise}. Also, the areas of applicability are different: the main advantage of the moment-based witness is that it is suitable in scenarios involving amplification. }
}


In the following, we first show the step-by-step proof for the proposed moment-based witness Eqs.~(\ref{eq:m}-\ref{eq:s2}). 
We will also discuss the extension of the witness in the multimode setting for the two extremes: independent modes and identical modes.
We furthermore provide equivalent formulations of the witness where the variance in the witness \eqref{eq:s2} is replaced by the appropriate values of the non-centered second moment \(\langle\hat N^2\rangle\), the second integrated intensity moment \(\langle W^2\rangle\), and the second-order correlation function \(g^{(2)}(0)\).


\subsection{Proof}\label{sec:proof}

The proof is structured as follows: we will take an arbitrary mixture, and show that each component's contribution to the second moment of the mixture can be reduced by replacing it with a specific \ernew{optimal displaced squeezed vacuum} without changing the contribution to the mean photon number. That is, we can reduce the photon number variance of any Gaussian mixture by replacing each component with the appropriate \ernew{displaced squeezed vacuum}. Then, we will show that the second moment of the photon number for such \ernew{optimal displaced squeezed vacua} is a convex function, and therefore, the minimum-variance Gaussian mixture for a fixed value of the mean photon number consists of a single \ernew{displaced squeezed vacuum state}, the photon numbers of which can be described by Eqs.~(\ref{eq:m}-\ref{eq:s2}). 

\paragraph{Parametrization of Gaussian states.} 
\ernew{
Any single Gaussian state can be constructed from a thermal state with mean photon number \(\overline n\) by first squeezing it along the \(x\) axis with squeezing parameter \(r\), then rotating it about the origin by an angle \(\varphi\), and finally displacing it by \((d_x, d_p)\). To make the construction unique, without the loss of generality, we will restrict ourselves to \(r \geqslant 0\) and \(\varphi \in [0, \pi)\). In this parameterization, the Wigner function of the resulting state is a two-dimensional Gaussian distribution centered at \((d_x, d_p)\) with the covariance matrix
\begin{align*}
    &C = \sigma^2\times\\
&\begin{bmatrix}
 e^{-2r}\cos ^2\varphi+e^{2r} \sin ^2\varphi & 
 -\sinh 2r \sin 2\varphi \\
-\sinh 2r \sin 2\varphi & 
 e^{-2r}\sin ^2\varphi+e^{2r} \cos ^2\varphi \\
\end{bmatrix},
\end{align*}
where we introduced the notation \(\sigma^2 \equiv (2\overline n +1)/4\) for the variance of the initial thermal state. We will further parameterize any Gaussian state through the tuple \((\sigma, r, \varphi, d_x, d_p)\).
Plugging this state into Eqs.~(\ref{eq:Wigner-mean}-\ref{eq:Wigner-var}), we obtain for the mean and the variance of the photon number of this state
\begin{align}
\label{eq:mean-Gau}
m &=  -\frac 12 + \tr C + d_x^2 + d_p^2, \\
\nonumber s^2 &= -\frac 1 4 - 4\sigma^4 + 2 (\tr C)^2  \\
\nonumber &+4\sigma^2 e^{-2r}(d_x\cos \varphi + d_p\sin \varphi )^2 \\
\label{eq:var-Gau}&+4\sigma^2 e^{2r}(-d_x\sin \varphi + d_p\cos \varphi )^2, 
\end{align}
where \(\tr C = 2\sigma^2 \cosh 2r\).
}

\begin{figure}[!t]
\centering
\begin{tikzpicture}
\draw[->,thick] (-1,0)--(6,0) node[right]{$x$};
\draw[->,thick] (0,-1.5)--(0,2) node[above]{$p$};
\draw[-] (1.5,0.1)--(1.5,-0.1) node[below]{$d_0$};
\draw[rotate around={60:(1.5,0)}] (1.5,0) ellipse (1.5cm and 0.5cm);
\draw[dashed, rotate around={60:(1.5,0)}] (0,0)--(3,0);
\draw[dashed, rotate around={60:(1.5,0)}] (1.5,-0.5)--(1.5,0.5);
\draw[red] (1.5,0) ellipse (0.5cm and 1.5cm);
\draw[green] (3.7,0) ellipse (0.25cm and 0.75cm);
\draw[-] (3.7,0.1)--(3.7,-0.1) node[below]{$d_2$};
\draw[purple] (5,0) ellipse (0.3cm and 0.625cm);
\draw[-] (5,0.1)--(5,-0.1) node[below]{$d_3$};
\draw (5, 0.9) node{(3)};
\draw (3.7, 1) node{(2)};
\draw (1.5, 1.75) node{(1)};
\draw (2.4, 1.55) node{(0)};
\end{tikzpicture}
\caption{\ernew{Steps of optimizing the photon number variance of a Gaussian state without changing the mean photon number. We start off with an arbitrary Gaussian state represented by ellipse (0), and first rotate it about its center \((d_0, 0)\) (1), and then increase its distance from the origin while keeping its squeezing parameter unchanged until we arrive at the displaced squeezed vacuum marked (2). State (3) represents the minimal photon number variance displaced squeezed vacuum given the mean photon number, and has the moments described in Eqs.~(\ref{eq:m}-\ref{eq:s2}).}\label{fig:orientation}}
\end{figure}
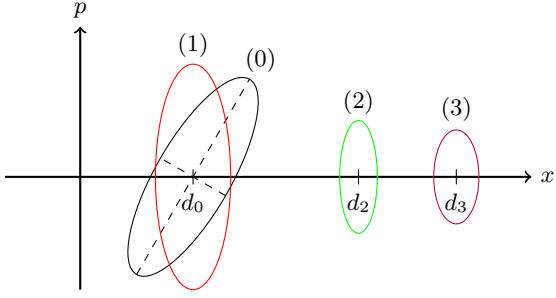

\paragraph{Orientation of the state.}
\ernew{Let us look at a single component of the mixture and choose a frame of reference in which its center lies on the \(x\) axis, in \(x = d_0\) (see ellipse (0) in Fig.~\ref{fig:orientation}). We will denote its parameters by \((\sigma_0, r_0, \varphi_0, d_0, 0)\). The corresponding photon number mean \(m_0\) and variance \(s_0^2\) from Eqs.~(\ref{eq:mean-Gau}-\ref{eq:var-Gau}) are 
\begin{align}
  \label{eq:proofa-m0} m_0 &= -\frac 12 + d_0^2 + \tr C_0,\\
  \nonumber s^2_0 &=-\frac 14 -4 \det \sigma_0^2 + 2(\tr C_0)^2 \\
  \label{eq:proofa-s2} &+ 4 d_0^2\sigma_0^2e^{-2r} + 2d_0^2\sigma_0^2\cosh2r_0 \sin^2\varphi_0.
\end{align}

Since the trace of \(C_0\) does not depend on the value of the rotation angle \(\varphi_0\), neither does \(m_0\). Therefore, the first goal is to find the value of \(\varphi\) that minimizes \(s^2_0\) while leaving \(r_0\) and \(d_0\) (and, as a consequence, \(m_0\)) unchanged. Looking at \eqref{eq:proofa-s2} and keeping in mind that we assumed \(r_0 \geqslant 0\) and \(\varphi_0\in [0, \pi)\), this is trivially achieved by setting the rotation angle to zero. In other words, reorienting the state such that the eigendirection of its covariance matrix corresponding to the smaller eigenvalue is aligned with the \(x\) axis results in a lower photon number variance while leaving the mean photon number unchanged.

This improved state is represented by the ellipse (1) in Fig.~\ref{fig:orientation}, and has the parameters \((\sigma_1 = \sigma_0, r_1 = r_0, \varphi_1 = 0, d_1 = d_0, 0)\). The corresponding covariance matrix and photon number moments are
\begin{equation}
    C_1 = \sigma_0^2\times
\begin{bmatrix}
 e^{-2r_0} & 
 0\\
0 & 
 e^{2r_0}
\end{bmatrix},
\end{equation}
and 
\begin{align}
\nonumber m_1 &= m_0 = -\frac 12 + d_0^2 + 2\sigma_0^2\cosh 2r_0 \\    
\nonumber s_1^2 &= -\frac 14 -4\sigma_0^4 + 2(\tr C_1)^2 + 4 d_0^2\sigma_0^2e^{-2r_0}\\
\label{eq:s21} &=  -\frac 14 -4\sigma_0^4 + 8\sigma_0^4\cosh^22r_0 + 4 d_0^2\sigma_0^2e^{-2r_0}.
\end{align}
}

\paragraph{Lower-variance \ernew{displaced squeezed vacuum}.}
\ernew{
Now that we have arrived at state (1), let us compare it with the \ernew{displaced squeezed vacuum} (2) in Fig.~\ref{fig:orientation} that has the same mean photon number and the same value of the squeezing parameter. 
That is, state (2) is represented by the tuple \((\sigma_2 = 1/2, r_2 = r_0, \varphi_2 = 0, d_2 \neq d_0, 0)\), and therefore has the covariance matrix
\[C_2 = \frac 14 \begin{bmatrix}
    e^{-2r_0}  & 0\\
    0 & e^{2r_0}
\end{bmatrix}.
\]
Using that the mean photon numbers are equal in states (1) and (2), we obtain
\begin{equation}
\label{eq:m1m2}
    d_0^2+ 2\sigma_0^2\cosh 2r_0 =d_2^2 + \frac{1}{2}\cosh 2r_0.
\end{equation}
Turning to variances, we have
\begin{equation}
\label{eq:s22}
s_2^2 = -\frac 12 + \frac12\cosh^22r_0 + d_2^2e^{-2r_0}.
\end{equation}
Therefore, using equations \eqref{eq:s21}, \eqref{eq:m1m2}, and \eqref{eq:s22}, we obtain
\begin{align*}
s_1^2 - s_2^2 &=
 e^{-2 r_0} \left(\sigma_0 ^2-\frac 1 4\right) \times\\
 &\left [4 d_0^2+e^{2 r_0} \left(-1+4 \sigma_0 ^2 \cosh 4 r_0+\sinh 4 r_0\right)\right].
\end{align*}
Each term in this equation is non-negative: \(\sigma_0 ^2-\frac 1 4\geqslant 0\), \(d_0^2 \geqslant 0\), \(-1+4\sigma_0^2\cosh 4r_0 \geqslant -1+4\sigma_0^2 \geqslant 0\), and \(\sinh 4r_0\geqslant 0\) since we use non-negative values only for the squeezing parameter. As a consequence, \(s_2^2 \leqslant s_1^2\), so we achieved a decrease in the variance without changing the mean photon number (\(m_2 = m_1 = m_0\)). The moments of state (2) are then
\begin{align*}
    m_2 &= -\frac 12 + d_2^2 + \frac 12 \cosh 2 r_0 = d_2^2 + \sinh^2r_0,  \\
    s_2^2 &= -\frac 12 + \frac12\cosh^22r_0 + d_2^2e^{-2r_0} = \frac {\sinh^22r_0}{2} + d_2^2e^{-2r_0}. 
\end{align*}

}

\paragraph{\ernew{Optimal displaced squeezed vacuum}.}
Now we will find the \ernew{displaced squeezed vacuum} with the minimal variance for a fixed value of the mean photon number \(m\). 
\ernew{This optimal squeezed vacuum is represented by ellipse (3) in Fig.~\ref{fig:orientation}, and has the parameters \((\sigma_3 = 1/2, r_3 = r\neq r_0, \varphi_3 = 0, d_3\neq d_2, 0)\). }
The optimization problem is then
\begin{align*}
    m_3 &= d_3^2 + \sinh^2r = m_0 = \text{fixed}\\
    s_3^2 &= \frac{\sinh^22r}{2}+e^{-2r}d_3^2 \\
    &= \frac{\sinh^22r}{2} + e^{-2r}(m_0-\sinh^2r) = \text{min.}
\end{align*}
This problem is straightforward to solve analytically by requiring \(\mathrm d s_3^2/\mathrm dr = 0\); this yields \(0 = 1 + e^{8r}-2e^{2r}-4e^{2r}m_0\). While the latter equation does not allow us to express the optimal \(r\) as a function of \(m_0\) in a closed form, it clearly works the other way around, yielding the parametric form in Eqs.~(\ref{eq:m}-\ref{eq:s2}). We can also express the optimal displacement as \(d^2_3 = e^{2 r} \left(e^{4 r}-1\right)/4\), which is non-negative for \(r\leqslant 0\), so there are no issues with non-physicality. 

\begin{figure}[htb]
\centering
\includegraphics[width=0.9\columnwidth]{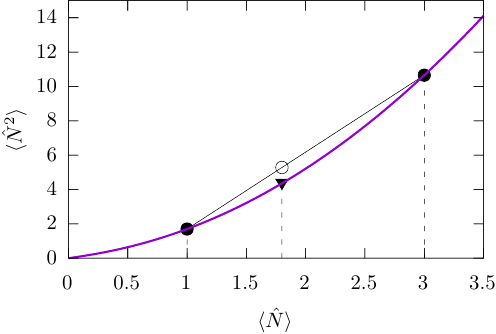}
\caption{\err{Behavior of the QNG witness} (\ref{eq:m}-\ref{eq:s2}) for a mixture of two \ernew{optimal displaced squeezed vacua}. \err{The purple line shows the non-centered second moment of an \ernew{optimal displaced squeezed vacuum} as a function of its mean photon number, that is, \(s^2_{\mathrm{NG}}(r)+m^2_{\mathrm{NG}}(r)\) as a function of \(m_{\mathrm{NG}}(r)\) (\(r\geqslant 0\)).} The mixture components are shown with black disks, with mean photon numbers one and three. Any state corresponding to their mixture lies in this representation on the thin black line connecting the two black disks. The black circle corresponds to a 60\%-40\% mixture of the initial states, which has a mean photon number of 1.8. 
\err{However, since the purple line represents a convex function, we obtain a lower second moment for a single \ernew{optimal displaced squeezed vacuum} with the same mean, 1.8 (black triangle).}
\label{fig:convexity}}
\end{figure}

\paragraph{Convexity of the second moment of the optimal displaced squeezed vacuum.}
So far, we have shown that any mixture of Gaussian states can be replaced by a mixture of optimal displaced squeezed vacua that have the same mean \(m\) but a smaller variance \(s^2\), or equivalently, a smaller non-centered second moment \(\langle \hat N^2\rangle\). Here, we will show that the second moment \(\langle \hat N^2\rangle\) of optimal displaced squeezed vacua is a convex function of the mean photon number \(m\). 
From the parametric form in Eqs.~(\ref{eq:m}-\ref{eq:s2}), we obtain
\[
\frac{\mathrm d^2 }{\mathrm d m_{\mathrm{NG}}^2}(s_{\mathrm{NG}}^2 + m_{\mathrm{NG}}^2) = 2-\frac{4}{3 e^{8 r}-1},
\]
which is non-negative for any \(r \geqslant 0\), implying convexity.
 Using this convexity, we can again apply Jensen's inequality to show that the second moment of the photon number of a single optimal displaced squeezed vacuum with a given mean \(m\) is smaller than that of the mixture of multiple \ernew{optimal displaced squeezed vacua} with the same mean photon number. Figure \ref{fig:convexity} demonstrates Jensen's inequality for two components in the mixture. For a multicomponent mixture, one can replace any pair of components by the \ernew{optimal displaced squeezed vacuum} having the same mean photon number as the average of the two components, and continue on with this procedure until only a single component remains. 

\ernew{That is, we have shown that for a fixed value \(m\) of the mean photon number, the mixture of Gaussian states that has the smallest value of \(\langle\hat N^2\rangle\) consists of a single Gaussian state, namely an optimal displaced squeezed vacuum. Note that, of course, a phase-randomized version of this state produces the same photon number moments.}

\subsection{Multimode case}
\ernew{Previously, we assumed} that the amplification process with a single-mode pump efficiently filters out a single output mode before the detection. However, we still discuss here some aspects for multimode light.
While tackling the general case of arbitrarily interdependent modes is not feasible, the two extremes of either all modes being independent or all modes being identical are instructive.

\paragraph{Independent modes.} 
\ernew{Considering \(M\) independent (uncorrelated) modes, 
and measuring the total number of photons (\(\hat N = \sum_{i = 1}^M\hat a^\dagger_i\hat a_i\)) corresponds to the sum of independent random variables, and therefore both the means and variances corresponding to the individual modes add up.} 
The mean and the variance per mode are then
\begin{equation*}
    \overline m = \frac 1 M \sum_{i = 1}^M m_i,\quad
    \overline{s^2} = \frac 1 M\sum_{i = 1}^M s_i^2.
\end{equation*}
\begin{figure}[htb]
\centering
\includegraphics[width=\linewidth]{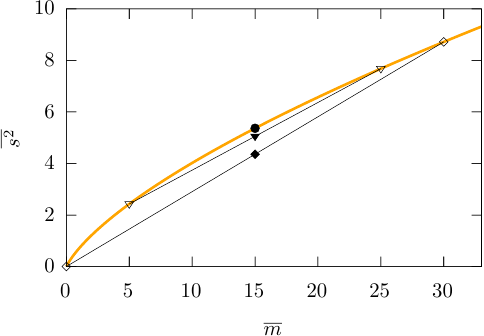}
\caption{\err{Behavior of the QNG witness} (\ref{eq:m}-\ref{eq:s2}) for two independent modes: number variance per mode and mean photon number per mode. The black disk corresponds to two identical optimally squeezed modes with \(m_1 = m_2 = 15\). The black filled triangle corresponds to the average from the two states denoted by the empty triangles (\(m_1 = 5\), \(m_2 = 25\)), the average photon number per mode is again 15; however, the variance is lower than in the identical case. The black filled diamond shows the average from \(m_1 = 0\) and \(m_2 = 30\) (empty diamonds): due to the convexity of the optimal variance curve, this is the setup that provides the lowest variance for \(\overline m = 15\).\label{fig:multi-indep}}
\end{figure}
Clearly, with \(m_i\) fixed, the minimal variance in each mode is attained by the appropriate displaced squeezed vacuum. 
\ernew{Taking two of these modes (with means $m_i$ and variances $s_i^2$, with $i\in\{1,2\}$), 
it can be shown} that replacing them by \ernew{optimal displaced squeezed vacua with} \(m_1' = m_1+m_2\) and \(m_2' = 0\), the resulting variance contribution will be smaller than the original, \(s_1'^2 + s_2'^2(=0) < s_1^2 + s_2^2\) (see Fig.~\ref{fig:multi-indep}, diamond vs.\ triangle). Continuing with this process pairwise, it follows that the lowest variance is attained by a single mode containing all the photons, and the remaining modes empty. Therefore, the previously attained single-mode QNG witness is applicable in the multimode case as-is if the modes are independent. 

Importantly, since beam splitters do not change the sum of the number of photons in a pair of modes, interdependent modes that are only dependent due to mixing on a network of beam splitters are, for our purposes, equivalent to the case of all modes being independent. Therefore, again, the non-Gaussianity witness is applicable in its single-mode form in such cases.

\begin{figure}[htb]
\centering
\includegraphics[width=0.9\columnwidth]{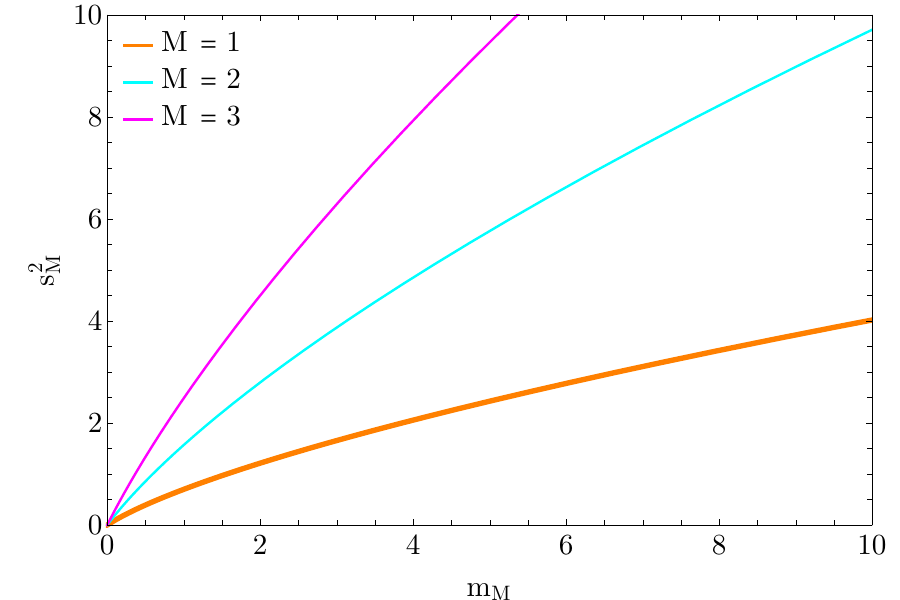}
\caption{Mean and variance of photon numbers for \ernew{optimal displaced squeezed vacua} in the case of \(M\) identical modes. \err{This illustrates that assuming \(M = 1\) provides the worst-case scenario in the identical multimode case, that is, if such a state can be represented by a point below the \(M = 1\) orange line, it also qualifies as QNG for any \(M > 1\).}\label{fig:multi-ident}}
\end{figure}

\paragraph{Identical modes.} As the other extreme, we may also consider \(M\) (\(M\geqslant 1\)) identical modes. Clearly, in this case we just multiply the photon numbers of a single mode by the number of modes, and therefore the moments become
\begin{equation*}
    m_M = M\cdot m_0,\quad
    s_M^2 = M^2 \cdot s_0^2,
\end{equation*}
with \(m_0\) and \(s_0^2\) denoting the mean and variance of the photon number of a single copy of the mode. Due to this simple form (both \(m_M\) and \(s^2_M\) strictly increasing with \(m_0\) and \(s_0^2\), respectively), the state that minimizes the variance is, again, an \ernew{optimal displaced squeezed vacuum}. Therefore, the minimum-variance curve can be given by
\[m_M = M\cdot m_{\mathrm{NG}}(r),\quad s_M^2 = M^2\cdot s^2_{\mathrm{NG}}(r).\]
Figure \ref{fig:multi-ident} shows that if the number of modes \(M\) is unknown, assuming \(M = 1\) is the worst-case scenario. That is, the single-mode witness provides a stricter criterion for non-Gaussianity than any higher-mode witness.

\subsection{Equivalent formulations}%


So far, for mathematical convenience, we worked mainly with the photon number mean and variance. In the following, we provide a few equivalent formulations of witness that may be handier depending on the scenario.

\paragraph{Formulation in terms of the non-centered second moment.} For any fixed mean photon number 
\[\ev{\hat N} = m_{\mathrm{NG}}(r) = \frac{e^{6r}}4 - \frac12 +\frac{e^{-2r}}{4}\]
parameterized by \(r \geqslant 0\), the non-centered second moment of the photon number is lower-bounded for any mixture of Gaussian states:
\begin{align*}
\ev{\hat N ^2} &\geqslant m^2_{\mathrm{NG}}(r)+s^2_{\mathrm{NG}}(r) \\
&= 
\frac{e^{12 r}}{16}
-\frac{e^{6 r}}{4}
+\frac{e^{4 r}}{2}
-\frac{1}{4}
-\frac{e^{-2 r}}{4}
+\frac{3 e^{-4 r}}{16},
\end{align*}
\err{see Fig.~\ref{fig:convexity}, purple line. Importantly, note that for mixtures of arbitrary states, it is the non-centered second moment that transforms linearly; that is, it is the convexity of this quantity that is relevant in the proof.}

\begin{figure}[!t]
\centering
\includegraphics[width=0.95\columnwidth]{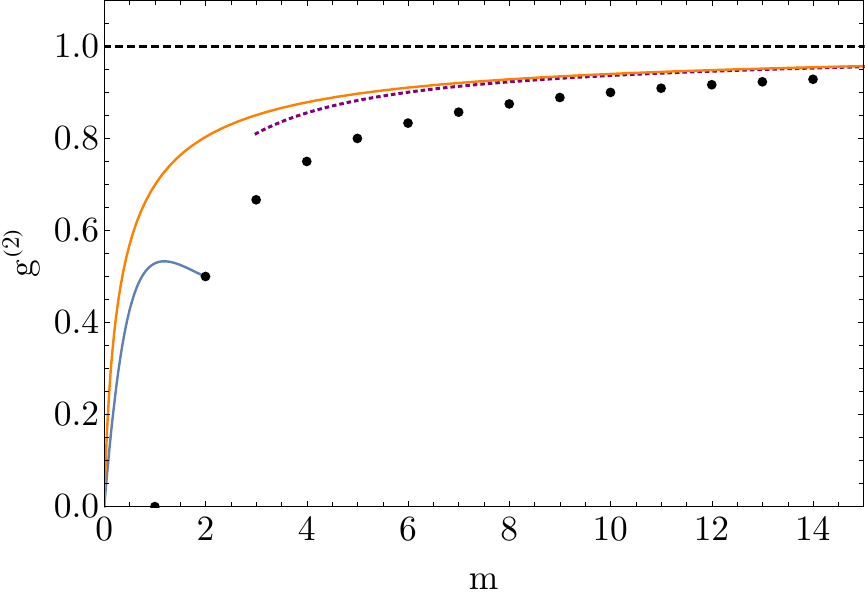}
\caption{Formulation of the QNG witness in terms of second-order correlation function and mean photon number. The black dashed line represents the witness for non-classicality, the orange line the witness of non-Gaussianity, whereas the blue line the transformed witness of quantum non-Gaussianity from (5) in \cite{Jezek2011}. The purple dotted line shows the asymptotic approximation of the orange line, \(1-(m+\frac12)^{-1}+3\cdot 2^{-5/3}(m+\frac12)^{-4/3}-(m+\frac12)^{-2}\). The black disks show the \(g^{(2)}\) values of Fock states \(n = 1,\ldots, 14\), \(1-n^{-1}\).
\label{fig:g2}}
\end{figure}

\paragraph{Formulation in terms of integrated intensity moments.}
\ernew{Advantageously, for the experiments with classical integrated-intensity detectors, one} can also describe the photon number distribution as a Poissonian random variable with a \ernew{random intensity parameter \(W\)}, the integrated intensity \cite{Mandel1995, Perina2017}, which corresponds to the detector's response over a finite time window. \ernew{In mathematical terms, this is a mixed Poisson distribution, and } we can express the relationship between the characteristic functions as \(C_N ( \xi) = C_W \left[(e^{i\xi}-1)/i\right]\) (see \cite{Mandel1995}, section 14.9). For example, a thermal number distribution can be obtained as a Poissonian with an exponentially distributed mean.   
The relationship between the moments of the integrated intensity and the photon number moments is simply \(\ev{W} = \ev {\hat N}\) and \(\ev{W^2} = \ev{\hat N^2}-\ev{\hat N}\). That is, for a fixed mean photon number \(m\), the second moment of the integrated intensity is a strictly increasing function of the second moment of the photon number, \ernew{so the second integrated intensity moment is minimal exactly when the second moment of the photon number is minimal}. The newly introduced QNG witness is therefore straightforward to express in terms of integrated intensity moments. For any fixed mean intensity
\[\ev{W} = m_{\mathrm{NG}}(r) = \frac{e^{6r}}4 - \frac12 +\frac{e^{-2r}}{4}\]
parameterized by \(r \geqslant 0\), the second moment of the integrated intensity is lower-bounded for any mixture of Gaussian states:
\begin{align*}
\ev{W ^2} &\geqslant m^2_{\mathrm{NG}}(r)+s^2_{\mathrm{NG}}(r)-m_{\mathrm{NG}}(r) \\
&= 
\frac{e^{12 r}}{16}
-\frac{e^{6 r}}{2}
+\frac{e^{4 r}}{2}
+\frac{1}{4}
-\frac{e^{-2 r}}{2}
+\frac{3 e^{-4 r}}{16}\\
&\equiv W^2_{\mathrm{NG}}(r).
\end{align*}
That is, if we have a pair of experimental averages \(\overline W\) and \(\overline{W^2}\), then \(\overline{W^2} < W^2_{\mathrm{NG}}\left[m_{\mathrm{NG}}^{-1}(\overline W)\right]\) is a witness of quantum non-Gaussianity (with \(m_{\mathrm{NG}}^{-1}\) denoting the inverse function of \(m_{\mathrm{NG}}\)).
\paragraph{Formulation in terms of the second-order correlation $g^{(2)}$.}
The second-order correlation function can be expressed from the mean and the variance of photon numbers as
\er{
\[g^{(2)} \equiv \frac{\ev{\normord{\hat N^2}}}{\ev{\hat N}^2} = \frac{\langle{W^2}\rangle}{\langle{ W}\rangle^2} \equiv 1+ \frac{s^2}{m^2}-\frac 1 m,\]
with the pair of colons denoting normal ordering.}
Clearly, for a fixed value of \(m\), \(g^{(2)}\) is linearly increasing with \(s^2\). As a consequence, for a fixed value of the mean photon number \(\ev{\hat N } = \ev{W} = m_{\mathrm{NG}}(r)\), the second-order correlation function is lower bounded for any mixture of Gaussian states:
\begin{align*}
g^{(2)}(r) &\geqslant   1+ \frac{s^2_{\mathrm{NG}}(r)}{m^2_{\mathrm{NG}}(r)}-\frac 1 {m_{\mathrm{NG}}(r)}\\
&=\frac{
e^{12 r}
+2 e^{10 r}
+3 e^{8 r}
-4 e^{6 r}
-3 e^{4 r}
-2 e^{2 r}
+3
}{\left(e^{6 r}+e^{4 r}+e^{2 r}-1\right)^2}\\
&\equiv g^{(2)}_{\mathrm{NG}}(r).
\end{align*}
\ernew{In other words, similarly to the previous instance, if the experimental value of \(g^{(2)}\) is lower than \(g^{(2)}_{\mathrm{NG}}\left[m_{\mathrm{NG}}^{-1}(\overline W)\right]\) (a point below the orange line in Fig.~\ref{fig:g2}), it is a witness of the quantum non-Gaussianity of the state.}

Figure \ref{fig:g2} shows the transformation of the probability-based witness with blue and the new witness with orange. The usual non-classicality witness of \(g^{(2)} < 1\) is shown with the dashed black line. 
\ernew{For large values of \(r\), we can use the approximation \(m_{\mathrm{NG}}(r) \approx e^{6r}/4-1/2\), and the QNG border in terms of \(g^{(2)}\) can be approximated from below
\begin{equation}
    g^{2}_{\mathrm{NG}} > 1 - \frac{1}{m+\frac12} + \frac{3}{2^{5/3}\left(m+\frac12\right)^{4/3}} -\frac{1}{\left(m+\frac12\right)^2}.
\end{equation}
This approximation is shown with the purple dotted line.
}
The black disks, on the other hand, correspond to the \(g^{(2)}\) values of pure Fock states, \(1-n^{-1}\), showing that \ernew{for not too large values of the mean photon number, there is enough distance between the orange line and the disks to certify the non-Gaussianity of states other than Fock states.}

\er{Note that similar results can be obtained for the Fano factor
\[\mathcal{F} \equiv \frac{\var (\hat N)}{\ev{\hat N}} = \frac{s^2}{m}.\]
Namely, the \(\mathcal{F}<1\) is a known nonclassicality witness. While using the Eqs.~(\ref{eq:m}-\ref{eq:s2}), one can get a similar witness for QNG using the Fano factor since if only mixtures of Gaussian states are considered,  
\[ \mathcal{F} \ge \frac{s^2_{\mathrm{NG}}(r)}{m_{\mathrm{NG}}(r)}\equiv \mathcal{F}_{\mathrm{NG}}(r).\]
}

\section{Examples of different measurement schemes}\label{sec:examples-schemes}

The moment-based non-Gaussianity witness is applicable in any situation where there is a reliable way to infer the values of \(m\) and \(s^2\) (or its alternatives). In what follows, we will show a few straightforward examples of doing that. Note that our aim here is not to solve a realistic experimental problem, but rather to show some examples where our witness could be applied. Therefore, our solution details only the ideal case, while for the noisy cases, a similar analysis could be applied as detailed in Sec.~\ref{sec:noise}.

These examples rely on the fact that we can express \ernew{the photon number moments \(m\equiv \langle \hat a^\dagger \hat a\rangle\) and \(s^2 \equiv \langle \hat a^\dagger \hat a\hat a^\dagger \hat a\rangle - m^2\)} by expanding the integral expressions at the beginning of Sec.~\ref{sec:results} to
\begin{align}
\label{eq:m-quad} m &= \ev{\hat x^2} + \ev{\hat p^2}-\frac 12,\\
\label{eq:s2-quad} s^2 &= \var \hat x^2 + \var \hat p^2 + 2\cov \left(\hat x^2, \hat p^2\right)-\frac 14,
\end{align}
where \(\var \hat O \equiv \ev{\hat O^2} - \ev{\hat O}^2\) and
\begin{align*}
    \cov \left(\hat x^2, \hat p^2\right) 
    &\equiv  
    \int x^2p^2W(x, p)\, \mathrm dx\mathrm dp - \ev{\hat x^2}\ev{\hat p^2} \\
    &= 
    \ev{\frac{\hat x^2\hat p^2 + \hat p^2 \hat x^2}{2} + \frac 18}- \ev{\hat x^2}\ev{\hat p^2}.
\end{align*}


\subsection{Witness from homodyning}\label{sec:homodyning}

\ernew{
Performing a homodyne measurement is the most straightforward approach to estimating quadrature moments. Looking at Eqs.~(\ref{eq:m-quad}-\ref{eq:s2-quad}), we need the moments \(\ev{\hat x^2}\), \(\ev{\hat p^2}\), \(\ev{\hat x^4}\), \(\ev{\hat p^4}\), and \(\int x^2p^2 W(x, p)\, \mathrm dx \mathrm dp\). While the first four of these are straightforward to obtain, \(\ev{\hat x^2\hat p^2}\) is not. However, by measuring in four directions: 0, \(\pi/4\), \(\pi/2\), and \(3\pi/4\), this cross term can also be obtained by showing that \(\ev{\hat q^4_{\pi/4}} + \ev{\hat q^4_{3\pi/4}} = \big(\ev{\hat q^4_0} + \ev{\hat q^4_{\pi/2}}\big) /{2} + 3\int x^2p^2 W(x, p)\,\mathrm d x \mathrm dp\), with \(\hat q_\varphi \equiv \hat x \cos \varphi + \hat p \sin \varphi\) denoting the quadrature in direction \(\varphi\). After straightforward manipulations, this leads to the following form for the photon number moments:
\begin{align}
\label{eq:homodyne-m}    m &= \frac12\left[\ev{\hat q^2_0} + \ev{\hat q^2_{\pi/4}} +  \ev{\hat q^2_{\pi/2}}+\ev{\hat q^2_{3\pi/4}} - 1\right],\\
\nonumber    s^2 &= 
\frac23\left[
\ev{\hat q^4_0} + \ev{\hat q^4_{\pi/4}} + \ev{\hat q^4_{\pi/2}} + \ev{\hat q^4_{3\pi/4}}\right] \\
\label{eq:homodyne-s2}&
- \frac14\left[\ev{\hat q^2_0} + \ev{\hat q^2_{\pi/4}} + \ev{\hat q^2_{\pi/2}} + \ev{\hat q^2_{3\pi/4}}\right]^2 -\frac 14.
\end{align}
So after measuring the quadrature moments in these four directions, one can estimate the photon number moments and check if they are a witness of quantum non-Gaussianity, that is, if the corresponding point \((m, s^2)\) lies below the orange line in Fig.~\ref{fig:squeezed-coherent}. Alternatively, we can convert the criterion from \(m\) and \(s^2\) to the average quadrature moments \(Q^2 \equiv \frac 14 \sum_{k = 0}^{3}\ev{\hat q_{k\pi/4}^2}\) and \(Q^4 \equiv \frac14\sum_{k = 0}^{3}\ev{\hat q_{k\pi/4}^4}\) as
\begin{align}
\label{eq:Q2}    Q^2_{\mathrm{NG}}(r) &= \frac{e^{6r}+e^{-2r}}8\\
\label{eq:Q4}
    Q^4_{\mathrm{NG}}(r) &= 
    \frac3{128}\left[
    e^{12r}+8e^{4r}- 4 +3e^{-4r}
    \right].
\end{align}
That is, we have witnessed QNG if for \(Q^2 = Q^2_{\mathrm{NG}}(r)\), the average of the fourth quadrature moments \(Q^4 < Q^4_{\mathrm{NG}}(r)\).
}

\subsubsection*{Phase-random local oscillator} 
\ernew{A broadly available detection scheme that involves amplification using a strong single-mode local oscillator is a homodyne measurement with a phase-random local oscillator.
Essentially, this approach} \ernew{assumes that the local oscillator is not phase-locked to the signal, so it} randomizes the phase of the state and subsequently measures a single quadrature. By changing variables, we obtain the mean photon number as
\begin{align*}
\ev{\hat N}&= \int \mathrm dx \mathrm dp\, \left(x^2 + p^2-\frac 12\right)W(x, p) \\
&= 
\int \mathrm dr \mathrm d\varphi\, \left(r^2-\frac 12\right)\cdot r W(r\cos \varphi, r\sin\varphi) \\
&=
2\pi \int_0^\infty \mathrm dr \, \left(r^2-\frac 12\right)\cdot r\int_{0}^{2\pi}\frac{\mathrm d \varphi}{2 \pi}W(r\cos \varphi, r\sin\varphi)\\
&= \int_0^\infty \mathrm dr \, \left(r^2-\frac 12\right)\tilde W(r),
\end{align*}
where
\[\tilde W(r) \equiv r\int_{0}^{2\pi}\mathrm d \varphi\, W(r\cos \varphi, r\sin\varphi) \]
is exactly the Wigner function of the state after phase-randomization in polar coordinates, including the Jacobi determinant
\ernew{
\[
\left|
\begin{matrix}
\frac{\partial x}{\partial r} & \frac{\partial x}{\partial \varphi}\\
\frac{\partial p}{\partial r} & \frac{\partial p}{\partial \varphi}
\end{matrix}
\right|=r.\]
}
Similarly,
\begin{align*}
\ev{\hat N^2}&= \int \mathrm dx \mathrm dp\, \left(x^2 + p^2\right)\left(x^2 + p^2- 1\right)W(x, p) \\
&= \int_0^\infty \mathrm dr \, r^2\left(r^2-1\right)\tilde W(r).
\end{align*}
Clearly, this also means that the phase-randomization step does not affect the photon number mean or variance. Using the symmetry of the Wigner function of this randomized state, Eqs.~(\ref{eq:homodyne-m}-\ref{eq:homodyne-s2}) simplify to
\begin{align}
    m &= 2 \ev{\hat q^2}-\frac 12, \\
    s^2 &= \frac 83 \ev{\hat q^4} - 4 \ev{\hat q^2}^2 - \frac 14
\end{align}
where \(\hat q\) denotes an arbitrary quadrature of the phase-randomized state. As a result, the non-Gaussianity witness derived in terms of \(m\) and \(s^2\) is easy to express via \ernew{\(\ev{\hat q^2}\) and \(\ev {\hat q^4}\): we have a witness of quantum non-Gaussianity if for \(\ev{\hat q^2} = Q^2_{\mathrm{NG}}(r)\), the fourth moment \(\ev {\hat q^4} < Q^4_{\mathrm{NG}}(r)\).
}
The quantities \(\ev{\hat q^2}\) and \(\ev{ \hat q^4}\) are, of course, straightforward to estimate from a finite sample of realizations of \(\hat q\) measured in the homodyne setup with a phase-random local oscillator. So all that remains is to check if the point determined by these empirical values lies below the parametric curve defined by Eqs.~(\ref{eq:Q2}-\ref{eq:Q4}). This witness (\ref{eq:Q2}-\ref{eq:Q4}) represents an alternative to the approach used in \cite{Jezek2012_Experimental}.

\subsection{Phase-sensitive amplification}
\ernew{
An alternative approach uses optical pre-amplification followed by integrated-intensity detection and analyzing photon-number moments estimated from them. \latest{This approach was first proposed and experimentally verified \cite{Kalash2025} to show the quantum non-Gaussianity of states without phase-dependence, specifically heralded single photons. In what follows, we extend that technique to arbitrary single-mode states: }
 using the amplified moments, we estimate the original values of \(m\) and \(s^2\) before amplification in order to be able to apply the witness given in (\ref{eq:m}-\ref{eq:s2}). 
}

The action of a phase-sensitive amplifier, assuming that it amplifies (anti-squeezes) the \(x\) quadrature, can be expressed as
\[
W'(x, p) =  W\left( x e^{-r}, pe^r\right)
\]
for some value \(r > 0\). This amounts to multiplying the \(x\) quadrature by \(e^r\) and the \(p\) quadrature by \(e^{-r}\). For high values of the parametric gain \(r\), the output intensity is then approximately proportional to \(\hat x^2\).  
\ernew{
Since Eqs.~(\ref{eq:homodyne-m}-\ref{eq:homodyne-s2}) only contain even moments,  parametric homodyning as described in \cite{Kalash2023, barakat2025simultaneous} is also applicable to verify quantum non-Gaussianity. This would require amplification in four directions, similarly to the standard homodyne case (Sec.~\ref{sec:homodyning}). 
}

\begin{figure}[!t]
    \centering
    \includegraphics[width=\columnwidth]{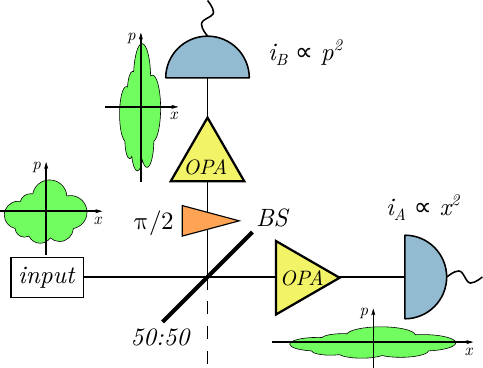}
    \caption{
    \ernew{Double parametric homodyne scheme. The input state is first split on a balanced beam splitter (BS). Then the state in the horizontal arm is strongly parametrically amplified (OPA) in the \(x\) quadrature direction, and the state in the vertical arm is first phase-shifted by \(\pi/2\) and then parametrically amplified by an identical amplifier (OPA). The integrated intensities are then measured in either arm, and are proportional to \(x^2\) and \(p^2\), respectively. }
    \label{fig:heterodyne}
    }
\end{figure}

Alternatively, note that that all terms in Eqs.~(\ref{eq:m-quad}-\ref{eq:s2-quad}) (\(\ev{\hat x^2}\), \(\ev{\hat p^2}\), \(\cov(\hat x^2, \hat p^2)\)) are directly available through a \ernew{double homodyne} setup, at the cost of additional noise. This noise can, however, be corrected for. For example, in the horizontal arm of Fig.~\ref{fig:heterodyne}, we have
\begin{align*}
    \hat x &= \frac 1{\sqrt 2}\hat x_0 + \frac 1{\sqrt 2}\hat x_{\mathrm{vac}}\\
    &\Downarrow\\
    \ev{\hat x^2} &= \frac 12 \ev{\hat x_0^2} + \frac 18\\
    \ev{\hat x^4} &= \frac14\left[\ev{\hat x_0^4} + \frac 32 \ev{\hat x_0^2}+\frac 18\right],
\end{align*}
where \(\hat x_{\mathrm{vac}}\) denotes the \(x\) quadrature of vacuum (\(\ev{\hat x_{\mathrm{vac}}} = \ev{\hat x_{\mathrm{vac}}^3} = 0\), \(\ev{\hat x_{\mathrm{vac}}^2} = 1/4\), \(\ev{\hat x_{\mathrm{vac}}^4}=3/16\)). The effect is identical when looking at the moments of \(\hat p\). The covariance of \(\hat x_0^2\) and \(\hat p_0^2\) is unaffected by the addition of vacuum, there is only a factor of \(1/4\) coming from the beam splitter. 
\ernew{
We can then plug in the quadrature moments that have been corrected for noise into Eqs.~(\ref{eq:m-quad}-\ref{eq:s2-quad}) to estimate \(m\) and \(s^2\), and finally check if the estimated point lies below the orange line in Fig.~\ref{fig:squeezed-coherent}.
}

\subsection{Phase-insensitive amplification}
A phase-insensitive amplifier amplifies the signal at the cost of adding noise from the idler mode. In the simplest form, its action can be described as \cite{Milburn1987}
\begin{align*}
    \hat x &= \sqrt{G}\cdot \hat x_s + \sqrt{G-1}\cdot\hat x_i\\
    \hat p &= \sqrt{G}\cdot\hat p_s - \sqrt{G-1}\cdot\hat p_i,
\end{align*}
for some gain \(G > 1\) with \((\hat x_{s/i}, \hat p_{s/i})\) denoting the signal/idler quadratures and \((\hat x,  \hat p)\) the output quadratures. Knowing the various quadrature moments of the idler mode, the amplified quadrature moments are straightforward to express and correct for. 

As the simplest example, let us consider the idler to be vacuum (\(\Rightarrow \ev{\hat x_i}=\ev{\hat p_i}=0\), \(\ev{\hat x_i^2}=\ev{\hat p_i^2}=1/4\), \(\ev{\hat x_i^4}=\ev{\hat p_i^4}=3/16\), \(\cov{(\hat x_i^2,\hat x_i)}= \cov({\hat p_i^2, \hat p_i})=0 \)). In this case, we can express the amplified photon number moments as
\begin{align}
\nonumber     M &= G\left(\ev{\hat x_s^2} + \ev{\hat p_s^2}\right) + \frac{G-1}{2}-\frac 12\\
    \label{eq:m-pia} &= m G  + G - 1,\\
\nonumber     S^2 &= G^2\left[\var \hat x^2 + \var \hat p^2 + 2\cov \left(\hat x^2, \hat p^2\right)\right]\\
\nonumber &+ G(G-1)\left(\ev{\hat x_s^2} + \ev{\hat p_s^2}\right)-\frac 14 \\
\label{eq:s2-pia} &= G^2s^2 + G(G-1)m + (3G^2-2G-1)/4,
\end{align}
with \(m\) and \(s^2\) denoting the initial photon number mean and variance of the signal.
\ernew{
Using Eqs.~(\ref{eq:m-pia}-\ref{eq:s2-pia}) and \(G\) known, we can estimate \(m\) and \(s^2\) from \(M\) and \(S^2\) and check whether these estimates meet the QNG witness in Eqs.~(\ref{eq:m}-\ref{eq:s2}). Of course, the same can be done based on integrated intensity moments, which amounts to replacing Eq.~\eqref{eq:s2-pia} by
\begin{align}
\nonumber     W^2 =& S^2+M^2-M\\
\nonumber =& G^2(s^2+m^2-m)\\
\label{eq:w2-pia}&+\frac{G-1}{4}
\left[ 7(G-1)+16Gm
\right]
\end{align}
Alternatively, we can substitute \(m\) and \(s^2\) in Eqs.~(\ref{eq:m-pia}-\ref{eq:s2-pia}) (or in Eqs.~\eqref{eq:m-pia} and \eqref{eq:w2-pia}) with \(m_{\mathrm{NG}}(r)\) and \(s^2_{\mathrm{NG}}(r)\) from Eqs.~(\ref{eq:m}-\ref{eq:s2}) to obtain a witness in terms of the amplified moments.  
}
\begin{figure}[ht]
    \centering
    \includegraphics[width=\linewidth]{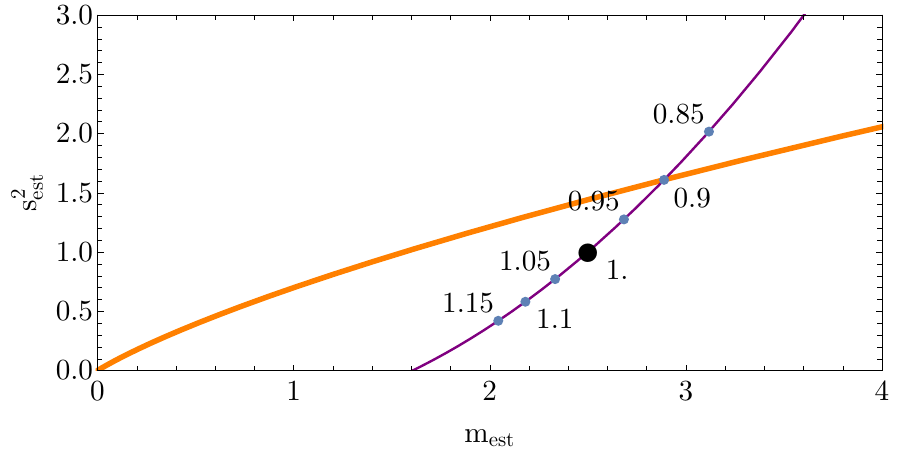}
    \caption{
    \ernew{Sensitivity of the corrected value of the photon number mean and variance to the estimated gain \(G_{\mathrm{est}}\). 
    Assuming that the true values are \(m = 2.5\), \(s^2 = 1\) (indicated with a black disk), and \(G = 100\), we calculated from Eqs.~(\ref{eq:m-pia}-\ref{eq:s2-pia}) the values \(M = 349\) and \(S^2 = 42200\). Then we inverted Eqs.~(\ref{eq:m-pia}-\ref{eq:s2-pia}) using different values of estimated gain \(G_{\mathrm{est}}\) (purple line). Specific values of \(G_{\mathrm{est}}/G\) are indicated with labeled blue disks along the purple line.
    The orange line shows the border of the moment-based QNG witness.}
    \label{fig:PIA}}
\end{figure}

\ernew{
In experiments, we know the value of \(G\) only up to a finite precision quantified by the error bars around \(G_{\mathrm{est}}\). 
We should therefore consider this finite precision in the estimation of the pre-amplification moments. 
Figure \ref{fig:PIA} shows how the estimated pre-amplification moments change when assuming different values for \(G_{\mathrm{est}}/G\). Clearly, by assuming a larger value for estimated gain \(G_{\mathrm{est}}\), both the variance and the mean decrease, moving deeper into the non-Gaussian region (possibly providing even non-physical values). Therefore, if we have a confidence interval \((G_{\min}, G_{\max})\), the conservative choice is to invert Eqs.~(\ref{eq:m-pia}-\ref{eq:s2-pia}) using \(G_{\min}\).}

\section{Resilience of quantum non-Gaussianity to loss and noise}\label{sec:resilience_to_loss_and_noise}
In the previous section, we looked at a few possible schemes where \(m\) and \(s^2\) may be estimated without photon-number resolving detectors. In the current section, we look at how the moment-based witness (\ref{eq:m}-\ref{eq:s2}) behaves assuming that these estimates are exact. 
\er{This allows us to explore the limits of the applicability of the moment-based QNG witness independently of the particulars of the experiment from which the estimates \(m_{\mathrm{est}}\) and \(s^2_{\mathrm{est}}\) were obtained. 
We can also look at it as the infinite sample size limit of the measurement, where \(m_{\mathrm{est}} = m\) and \(s^2_{\mathrm{est}} = s^2\).}

 
 
 To characterize the non-Gaussianity of a state, we will use the quantum non-Gaussianity depth (QNG depth) as a figure of merit \cite{Straka2014}. This is operationally the maximal loss (\ernew{or equivalently, the minimal transmittance $\eta_{min}$}) after which the attenuated state still shows quantum non-Gaussianity . Note that this number strongly depends on which QNG witness we use. \ernew{So, for the sake of simplicity, we will again} compare our new witness to the witness based on Fock-0 and Fock-1 probabilities \cite{Jezek2011}. 

\ernew{In the following, we first look at the quantum non-Gaussianity depth of Fock states and photon-added thermal states as examples. We will further examine how loss and additive noise affect the witness and how it may be corrected for.}

\subsection{QNG depth of Fock states}
As the simplest demonstration of the applicability of the witness (\ref{eq:m}-\ref{eq:s2}), we calculate how much signal loss a Fock state can tolerate in order for the witness to still recognize its non-Gaussianity. It is straightforward to show that if a Fock state \(\left|n\right\rangle\) passes through a beam splitter with transmissivity \(\eta\) 
its photon number mean and variance become
\begin{equation}\label{eq:Fock}
\begin{split}
m &= \eta n , \\
s^2 &= \eta(1-\eta) n.
\end{split}
\end{equation}

\begin{figure}[!t]
\centering
\begin{tikzpicture}
\node[anchor=north east, inner sep=0] (top-left) at (0,0){\includegraphics[width=\linewidth]{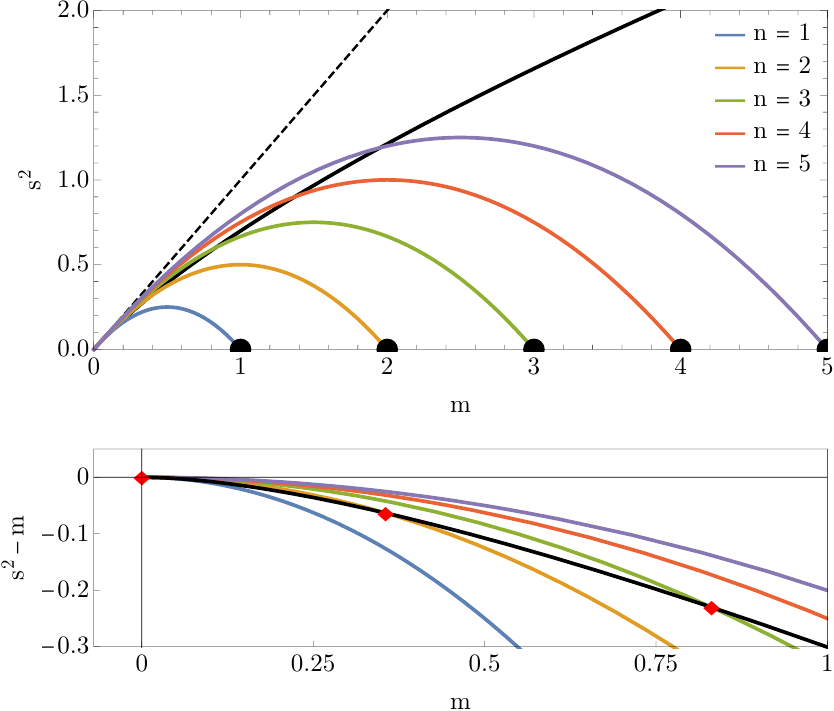}};
\node[inner ysep=8, inner xsep = 29, anchor=north west] (a) at (top-left.north west){\footnotesize(a)};
\node[inner ysep=20, inner xsep = 29, anchor=south west] (a) at (top-left.south west){\footnotesize (b)};
\end{tikzpicture}
\caption{QNG depth of Fock states without added noise. (a) The photon number mean and variance of lossy Fock states 1-5, the newly obtained QNG witness is shown with the black solid line. The QNG depths are given by the intersections of the colored curves with the black line. The black disks indicate the lossless, \(\eta = 1\) states. The dashed black line shows the non-classicality border \(s^2 = m\). (b) \ernew{The same lines depicted via \(s^2-m\)} 
as a function of the mean. The numerically calculated intersection points for Fock states 1-3 are indicated with red diamonds.
\label{fig:QNG1}}
\end{figure}

For any set of values \(n\) and \(\eta\) 
it is easy to check whether \((m, s^2)\) lies below or above the border of the non-Gaussianity witness given by Eqs.~(\ref{eq:m}-\ref{eq:s2}). 
Fig.~\ref{fig:QNG1}(a) shows how the moments of Fock states 1-5 change as they pass through a beam splitter with different values of the transmissivity \(\eta\in[0,1]\). For \(\eta = 1\), each curve starts at \((n, 0)\) (black disks), and arrives in the origin when \(\eta\) is decreased to zero. For \(n > 1\), each curve intersects with the QNG border for some value \(\eta^* > 0\) (which defines the QNG depth), and this \(\eta^*\) increases with \(n\). This means that, expectedly, higher Fock states can tolerate less loss, that is, their QNG depth is smaller.
\begin{table}[h]
\centering
\caption{QNG depth \ernew{\(1-\eta_{\min}\)} of different Fock states via the probability-based witness and the new, moment-based witness.\label{table:QNGdepth}}
\begingroup
\setlength{\tabcolsep}{10pt}
\begin{tabular}{ccc}
\toprule
\multirow{ 2}{*}{\(n\vphantom{3^{3^{3}}}\)}& \multicolumn{2}{c}{\(1-\eta_{\min}\)} \\
\cmidrule(lr){2-3}
 & Prob.-based & Moment-based\\
\midrule
 1 & 1 & 1 \\
 2 & 0.63 & 0.82 \\
 3 & 0.51 & 0.72 \\
 4 & 0.46 & 0.66 \\
 5 & 0.42 & 0.61 \\
 \bottomrule
\end{tabular}
\endgroup
\end{table}

Table \ref{table:QNGdepth} compares the QNG depths using the probability-based witness (\ernew{using that \(p_0 = (1-\eta)^n\) and \(p_1 = n\eta(1-\eta)^{n-1}\) for lossy Fock states}) 
and the moment-based witness. For the Fock-1 case, we have a mixture of vacuum and single photon, which is always non-Gaussian. For Fock-2 and higher, the new witness (\ref{eq:m}-\ref{eq:s2}) provides a considerably larger depth (smaller \(\eta_{\min}\)) for all cases without a need to specify photon-number probabilities. Note that for the probability-based witness, this assumes that we can estimate both \(p_0\) and \(p_1\) quite accurately, even if they are both close to zero, which might be difficult in practice.

\begin{figure}[!t]
\includegraphics[width=.8\columnwidth]{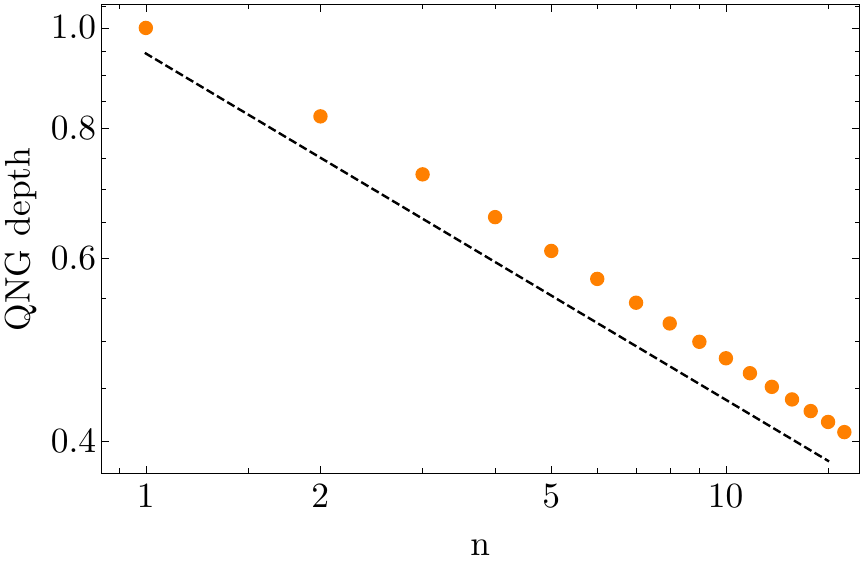}
\caption{The numerically calculated QNG depths (defined by the witness Eqs.~(\ref{eq:m}-\ref{eq:s2})) of Fock states 1-15 as a function of \(n\) (orange points) on a doubly logarithmic scale and the asymptotic slope of -1/3 indicated for reference.
\label{fig:QNG_large}}
\end{figure}

Let us note that our witness extends the applicability range from microscopic states ($m\ll 1$) to mesoscopic states ($m$ could be well above 1), but in principle it can be applied to any state. For large values of the squeezing parameter \(r\), Eqs.~(\ref{eq:m}-\ref{eq:s2}) simplify to
\begin{equation}
\label{eq:NGasymp}
\begin{split}
m_{\mathrm{NG}}(r) &\approx \frac{e^{6r}}{4}\\
s^2_{\mathrm{NG}}(r) &\approx \frac{3e^{4r}}{8} = \frac38 (4m)^{2/3},
\end{split}
\end{equation}
and therefore the equation for QNG depth can be solved asymptotically, and yields \(1-\eta_{\min} = \frac{3}{8} 4^{2/3} n^{-1/3}\) (see Fig.\ \ref{fig:QNG_large}, black dashed line indicating the asymptotic slope). We can see that this analytic approximation is close to the numerically calculated values for large Fock states (orange points). 
Note that the compared $(p_0, p_1)$-based witness is most suited for low mean photon numbers. A similar extension of range for mesoscopic states could be achieved also to the probability-based witness, but that needs a more sophisiticated setup 
\er{with multiple single photon detectors \cite{Lachman2016, Straka2018} and/or a more complex witness involving only numerically attainable critical values \cite{Fiurasek2025}.}


\subsection{QNG depth of photon-added thermal states}

The photon number distribution of a \(k\)-photon-added thermal state with (initial) mean photon number \(\overline n\) can be given as \cite{Barnett2018}:
\begin{equation}
   p_{+k}(n) = \frac{\overline n^{n-k}}{(1+\overline n)^{n+1}}\binom{n}{k};\quad n \geqslant k,
\end{equation}
which\ernew{, in mathematical terms, is a random variable obtained by adding \(k\) to a negative binomial distribution with parameters \(k+1\) and \((1+\overline n)^{-1}\) (i.e., its distributed as \(\mathrm{NB}(k+1, 1/(1+\overline n))+k\)).}
Note that this describes the case of adding $k$ photons realistically to a thermal state, meaning that $\overline n=0$ is equivalent to our previous case of Fock states. 

The photon number mean and variance are therefore
\begin{align*}
    m_{+k}(\overline n) &= k+(k+1)\overline n\\
    s^2_{+k}(\overline n) &= (k + 1)\overline n(\overline n+1).
\end{align*}
After passing through a beam splitter with transmittance \(\eta\), these moments change to
\begin{equation}\label{eq:photon_added}
\begin{split}
    m_{+k}(\overline n, \eta) &= \eta\left[ k+(k+1)\overline n\right]\\
    s^2_{+k}(\overline n, \eta) &= \eta^2(k + 1)\overline n(\overline n+1)\\
    &+\eta(1-\eta)\left[k+(k+1)\overline n\right].
\end{split}
\end{equation}

\begin{figure}[!t]
\centering
\begin{tikzpicture}
\node[anchor=north east, inner sep=0] (top-left) at (0,0){\includegraphics[width=\linewidth]{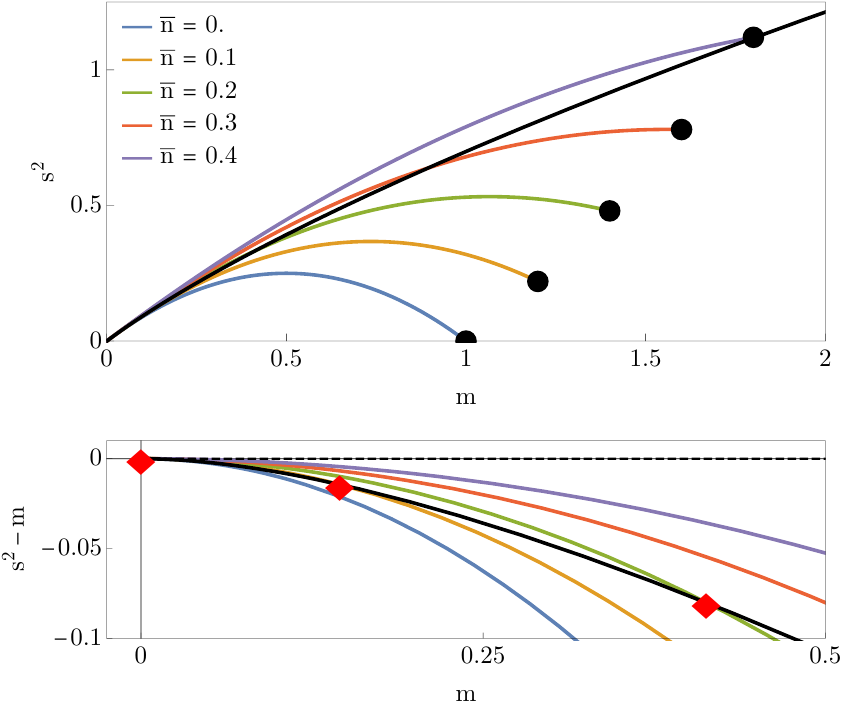}};
\node[inner ysep=20, inner xsep = 41, anchor=south west] (a) at (top-left.west){\footnotesize(a)};
\node[inner ysep=25, inner xsep = 41, anchor=south west] (b) at (top-left.south west){\footnotesize (b)};
\end{tikzpicture}
\caption{QNG of single photon-added thermal states. (a) Photon number mean and variance of single photon added thermal states with initial mean photon number \(\overline n\) between 0 and 0.4. The black disks correspond to the lossless, \(\eta = 1\) case. (b) \(s^2-m\) for a narrower range of mean values. The red diamonds indicate where the curves corresponding to \(\overline n= 0, 0.1\) and 0.2 intersect the border of the moment-based QNG witness.\label{fig:PATS}}
\end{figure}

Figure~\ref{fig:PATS} shows the photon number mean and variance of single-photon-added thermal states as \(\overline n\) is increased from $0$ to $0.4$. We can see that the $\overline n=0$ case coincides with a lossy single photon, which is always in the QNG regime for \(\eta > 0\). If we increase the size of the initial thermal state, the variance of the lossless state (\(\eta = 1\), black disks) increases, until at \(\overline n \approx 0.4\), it reaches the QNG border. This means that the moment-based witness cannot show the non-Gaussianity of single-photon-added thermal states whose initial mean photon number \(\overline n\) is over 0.4.



\begin{table}[htb]
\centering
\caption{QNG depths \ernew{\(1-\eta_{\min}\)} of \(k\)-photon added thermal states with initial mean photon number \(\overline n\) using the probability- (P) and the moment-based (M) witnesses.\label{table:QNGdepth2}}
\begingroup
\setlength{\tabcolsep}{8pt}
\begin{tabular}{ccccccc}
\toprule
\multirow{ 2}{*}{\(\overline n\vphantom{3^{3^{3}}}\)}
& \multicolumn{2}{c}{\(k = 1\)} 
& \multicolumn{2}{c}{\(k = 2\)} 
& \multicolumn{2}{c}{\(k = 3\)}\\
\cmidrule(lr){2-3}
\cmidrule(lr){4-5}
\cmidrule(lr){6-7}
& P& M & P&M&P&M\\
\midrule
 
 0  & 
 1 & 1 &
0.63 &  0.82 &
 0.51 &0.72   \\
 0.1 & 
 0.76 & 0.88 & 
 0.54 &0.71 & 
 0.45 &0.61   \\
0.2 & 
0.62 & 0.71 &  
0.47 &0.56 & 
 0.41& 0.47  \\
0.3 & 
0.52 & 0.44 & 
0.42 & 0.36 & 
0.38 & 0.29   \\
0.4 & 
0.45 & 0 & 
0.38 & 0.06 & 
0.35 & 0.02\\
 \bottomrule
\end{tabular}
\endgroup
\end{table}

The numerically calculated values of QNG depth for both the probability-based and the moment-based witnesses are collected in Table \ref{table:QNGdepth2}. We can see that if we add more and more photons, then the QNG depth mostly decreases. That is, by adding more photons, we do not obtain a more non-Gaussian state. If that is our main goal, then adding only a single photon is more efficient. 

On the other hand, if we increase the initial mean photon number $\overline n$ of the thermal state, the QNG depth also decreases. This is again expected behavior, as the contribution of the classical thermal state will be higher compared to the added Fock state. For low values of $\overline n$, the moment-based method shows again a higher QNG depth compared to the probability-based witness. This, however, changes for larger values of $\overline n$, where the moment-based method fails to work around $\overline n \approx 0.4$ as the same moments can be increasingly well approximated with Gaussian distributions. In contrast, the probability-based method works even for larger thermal states. This is because for \(\eta = 1\), the probability of zero photons is always zero for any photon-added thermal state, which automatically implies non-Gaussianity. However, in a realistic experiment with different noises and imperfect detection, the property of $p_0=0$ is difficult to achieve. 

\err{\subsection{Loss and noise correction}\label{sec:noise}}


\ernew{As an extension to the previously investigated cases of Fock states \eqref{eq:Fock} and photon-added thermal states \eqref{eq:photon_added}, let us look at a general state with initial photon number moments \(m_0\) and \(s_0^2\) and assume that it suffers loss that can be described with the aggregate transmissivity \(\eta_{\mathrm{agg}}\) and an (aggregated) independent additive noise with mean $\mnoise$ and variance $\varnoise$. The altered moments can be given as 
\begin{equation}\label{eq:Fock_noisy}
\begin{split}
m &= \eta_{\text{agg}} \cdot m_0 + \mnoise , \\
s^{2} &= \eta_{\text{agg}}(1-\eta_{\text{agg}}) m_0 + \eta^2 s_0^2 + \varnoise.
\end{split}
\end{equation}}
The added loss $\eta_{\text{loss}}$ has the same effect as the attenuation for QNG depth $\eta$, the two of them will have a combined attenuation of $\eta_{\text{agg}}=\eta_{\text{loss}}\times\eta$. We have two choices on how to handle this situation:
\begin{enumerate}
\item \textbf{Accepting loss:} We check the non-Gaussianity of the lossy state, in which case the QNG depth will be simply smaller (\(\eta_{\min}' = \eta_{\min}/\eta_{\text{loss}}\)).
\item \textbf{Removing loss:} If we know the value of loss $\eta_{\text{loss}}$, we may simply correct the moments of our state for the given loss. In this case, the QNG depth does not change. 
\end{enumerate}

Which case we decide on depends on the actual situation. In general, removing the loss is applicable if we have a precise understanding of the loss (e.g., correcting for detection inefficiency), so that we do not overcompensate. Note that if we want to be precise, we should obtain a confidence interval for the loss and take the lower bound to remain on the safe side. 

The situation for the noises is very similar: if one knows both \(\mnoise\) and \(\varnoise\) exactly (by measuring the properties of the noise separately), they can be subtracted from the empirical values of \(m\) and \(s^2\) to remove their bias. Our method of using moments is extremely useful here, as for additive noises, the mean and the variance simply add up, and so they can be very easily corrected for noise. In contrast, if we are interested in the probability distribution, then an additive noise means a convolution, the inversion of which is numerically very difficult. So again, if we can characterize the imperfection precisely, we can remove its effect entirely. 
\ernew{Note that in practice, different noises can have a different effect in Eq.~\eqref{eq:Fock_noisy}, e.g., in the case of amplification and a Gaussian additive noise, the bias $\mnoise$ will come from the variance for pre-amplification noise, and will come from the mean for post-amplification noise.}

\begin{figure}[!t]
    \centering
    \includegraphics[width=\linewidth]{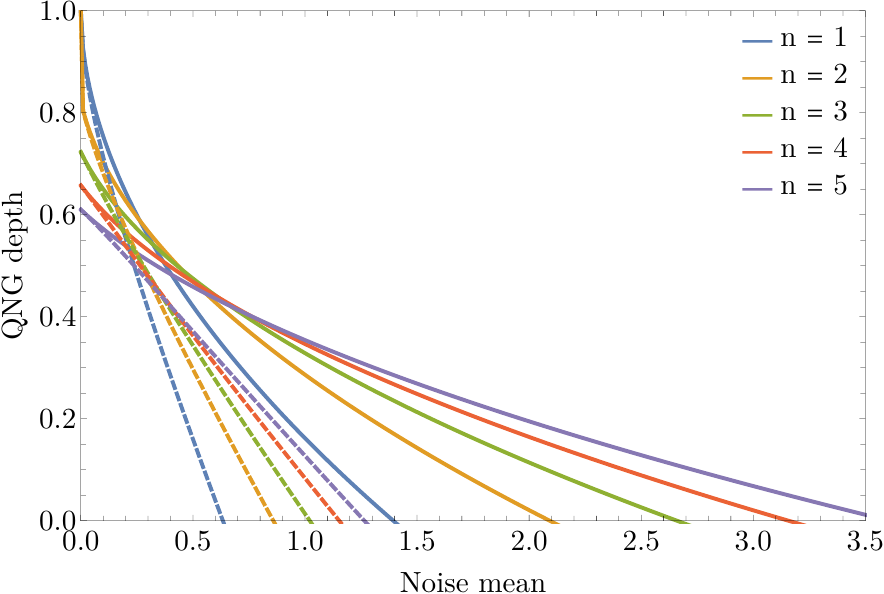}
    \caption{QNG depth of Fock states one to five assuming an independent noise added during detection. Solid lines: Poissonian noise; dashed lines: thermal noise. The dashed and solid lines of the same color correspond to the same Fock state \ernew{(see legend)}.}
    \label{fig:QNGwPOI}
\end{figure}

For the sake of completeness, let us also check what happens if we cannot remove the noise. In this case, the QNG depth decreases, but not as trivially as for losses. 
\ernew{Let us look at a numerical example of Fock states with two different types of noises (Fig.~\ref{fig:QNGwPOI}). This means we have the following values in (\ref{eq:Fock_noisy}): $m_0=n$ and $s_0=0$.} 
In the first case (solid lines), the noise is Poissonian with mean \(\overline n\), so we have \(\mnoise = \varnoise = \overline n\). In the second case (dashed line), the noise is thermal with mean \(\overline n\), so then \(\mnoise = \overline n\) and \(\varnoise = \overline n(\overline n +1)\). 

The behavior of the QNG depth is very similar in both cases: more added noise decreases the QNG depth of Fock states, and the higher the Fock state, the less sensitive its QNG depth is to this bias. The difference is that for the same noise mean, the Poissonian noise gives a significantly higher QNG depth, as it has a smaller variance than the thermal noise for the same mean. Note that the tolerable noise in either case has the same magnitude as the signal, which gives a good applicability for practical scenarios. If we have an even worse signal-to-noise ratio, the earlier discussed option of subtracting the estimated values of \(\mnoise\) and \(\varnoise\) helps to reduce the disruptive effect of the added noise.

\section{Conclusion and discussion}

Non-Gaussian states are important for different quantum applications. One may characterize such states by performing a complete tomography, but if we are interested only in their non-Gaussianity, providing a non-Gaussianity witness is a sufficient and, in general, simpler task. The standard method to achieve this witness in the literature is to perform a coincidence measurement, which can characterize the probability distribution of the unknown state for low photon numbers. A witness based on zero- and single-photon probabilities \cite{Jezek2011} can then be applied to establish \ernew{the quantum non-Gaussianity of a state}: for a fixed value of zero-photon probability \(p_0\), there is a maximal value of the single-photon probability \(p_1\) that any mixture of Gaussian states might reach. If the experimental value of \(p_1\) exceeds this limit, one can certify quantum non-Gaussianity.
This approach works well for specific tasks, but it needs single-photon detectors, which limits its applicability. 

In our work, we introduced a new witness for the non-Gaussianity of quantum states that relies only on measuring the mean and the variance of photon numbers. Similarly to its Fock probability based counterpart, we obtained a parametric curve Eqs.~(\ref{eq:m}-\ref{eq:s2}) (Fig.\ \ref{fig:squeezed-coherent}, orange line), which provides a lower limit on the photon number variance achievable by mixtures of Gaussian states given a fixed value of the mean. Thus, any state whose photon number variance lies below this limit has to be non-Gaussian. By converting the probability-based witness to photon number mean and variance, we can compare its behavior to the new witness: their borders coincide for low values of the mean photon number, but the new witness significantly relaxes the QNG requirement for larger values. 
\ernew{We provided alternative formulations of the witness, among others, we expressed it} in terms of the mean and the second-order correlation $g^{(2)}$ (Fig.\ \ref{fig:g2}, orange line), which provides a simple QNG witness similar to the $g^{(2)}<1$ witness of non-classicality (dashed line). \er{Note that a current, independent approach came to our attention, which uses correlation functions to characterize the non-Gaussianity of states \cite{Hotter2025}. Similarly to the standard approach, they also use single-photon detectors, importantly, our witness does not need single photon detectors, so} it is applicable in any scenario that allows one to infer the first two photon number moments of the state. 
We also examined the multimode case for the two extremes: independent modes and identical modes. In either case, the single-mode witness is applicable.

The new witness is applicable in any context where we can reliably estimate the photon number mean and variance, and we provided a few of such experimental scenarios. In general, the estimation of the mean for a dim state is not straightforward, but there are multiple methods that can obtain it with high precision. One of the methods is to apply a local oscillator and perform a homodyne measurement, which could characterize the state completely. We showed that for our witness, a non-phase-locked local oscillator is enough, which provides a homodyning in a phase-randomized way. Another common method to measure the mean is to amplify the signal and measure the intensity of the amplified state. We showed that in both the phase-insensitive and phase-sensitive cases, we may apply our QNG witness with minimal modifications. Note that we provided our examples for an ideal case as a proof-of-principle demonstration. However, since we showed that practical losses and noises are straightforward to deal with, we expect our witness to be applicable to real (mesoscopic and/or multimode) experimental setups (such as \cite{Kalash2025, Nechushtan2025}), extending the characterization of non-Gaussianity well above the currently achievable limits.

Finally, assuming that we have obtained accurate estimates of the photon number mean and variance of the input state, we analyzed this witness for a few simple states as an example. For this purpose, we used the notion of quantum non-Gaussianity (QNG) depth, which is defined as the maximum value of loss \ernew{(\(1-\eta_{\min}\))} a given state can suffer and still be witnessed as non-Gaussian. We investigated Fock states and photon-added thermal states, and in both cases, we achieved a decent QNG depth, which is better than the standard \ernew{\((p_0, p_1)\)} witness in practical scenarios. As a further advantage, due to the favorable properties of the statistical mean and variance, any loss and additive intensity noise can be very easily dealt with. To obtain the probability distribution after such imperfections, we need to deconvolve the noise, which is numerically a difficult task. In contrast, noise only modifies the moments linearly, so by accurately characterizing the magnitude of loss and noise precisely, we can easily minimize their effect. 
\ernew{
Note that knowing the full Wigner function or, at least the photon number probabilities for higher values of the photon number, provides, of course, more information than merely \((m, s^2)\) or the probabilites of \((p_0, p_1)\) from the compared witness. However, estimating the Wigner function or higher photon number probabilities is a more complex task (technically and/or witness-wise) than the estimation of the first two photon number moments. So another advantage of the moment-based witness is exactly that it is applicable to brighter states without added complexity.
}

In summary, we provided a new witness for quantum non-Gaussianity based on the photon number mean and variance, which is particularly suited for experiments involving high-gain amplification. Other than providing an extension of the applicability of QNG witnesses to brighter states, as a further advantage, it also provides straightforward ways to correct for loss and noise. \er{As an outlook, this approach also conveniently lends itself to an extension toward higher-order quantum non-Gaussianity as a counterpart of \cite{Straka2018, Lachman2019_NGhierarchy}.}

\er{\section*{Acknowledgments}  This research was funded within the QuantERA II Programme (project SPARQL), which has received funding from the European Union’s Horizon 2020 research and innovation programme under Grant Agreement No.~101017733. E.~R.\ and L.~R.\ also acknowledge the grant 23-06224S of the Czech Science Foundation. R.~F. acknowledges funding from the Horizon Europe Research and Innovation Actions under Grant Agreement no. 101080173 (CLUSTEC) \latest{and
the QuantERA project CLUSSTAR (8C24003) of EU and MEYS, Czech Republic.  Project CLUSSTAR has received funding from the European
Union’s Horizon 2020 Research and Innovation Programme under Grant
Agreement No.731473 and 101017733 (QuantERA).} R.~F.\ acknowledges discussions with Ch.~Hotter and A.~S.\ Sørensen, who independently derived quantum non-Gaussian criteria for correlation functions.}

\bibliography{moment_NG.bib}{}
\appendix

\section{Converting the Fock probability-based witness to photon number mean and variance}\label{sec:conversion}
\begin{figure}[!t]
\centering
\vspace*{5mm}
\includegraphics[width=0.8\columnwidth]{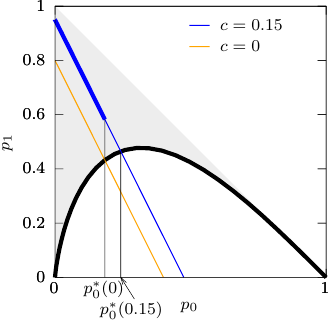}
\caption{
\ernew{
Illustration of converting the Fock probability based QNG witness Eq.~\eqref{eq:p0p1-criterion} to a witness in terms of \(m\) and \(s^2\). The probability-based witness identifies states as non-Gaussian if they lie above the thick black line, in the grey area (above the grey area, \(p_0 + p_1 > 1\), the states are non-physical). 
For a fixed value of the mean photon number \(m\) and \(c = 0.15\), the potential states lie on the blue line (see Eq.~\eqref{eq:conversion}). Therefore, in this example, we have a non-Gaussianity witness for \(p_0 < p_0^*(0.15)\). However, we do not actually know the value of \(c\). Nevertheless, assuming \(c = 0\) (orange line) and check the non-Gaussianity criteria for that, \(p_0 < p_0^*(0)\), makes sure that we are in the grey area, since then \(p_0<p_0^*(0) \leqslant p_0^*(0.15)\) (i.e., the state will be in thicker part of the blue line).
}
\label{fig:NGconversion}}
\end{figure}

The non-Gaussianity witness in terms of zero- and single-photon probabilities from \cite{Jezek2011} is 
\begin{equation}
    \label{eq:p0p1-criterion}
    \begin{split}
        \tilde p_0 (r) &=\frac{\exp\{-\frac{1}{4}(e^{4r}-1)(1-\tanh{r})\}}{\cosh{r}}
        ,\\
        \tilde p_1 (r) &=\frac{(e^{4r}-1)\exp\{-\frac{1}{4}(e^{4r}-1)(1-\tanh{r})\}}{4\cosh^3{r}}
    .
    \end{split}
    \end{equation}
with \(r \geqslant 0\) (see black line in Fig.~\ref{fig:NGconversion}). This means that for any fixed value of vacuum probability \(p_0 = \tilde p_0(r)\), the value of the single-photon probability cannot exceed \(\tilde p_1(r)\) for any mixture of Gaussian states. Therefore, given an experimental pair of values \((p_0, p_1)\) that lies above the black line in Fig.~\ref{fig:NGconversion} \ernew{(in the grey area)}, one can certify non-Gaussianity. The witness does not say anything about the non-Gaussianity of states that lie below this black line.

After straightforward manipulations of the definitions, we can calculate the photon number mean and variance as
\begin{align}\label{eq:ms}
m &\equiv \sum_{n = 0}^\infty np_n = 2 -  2p_0 - p_1 + 
\sum_{n=2}^{\infty}p_n(n-2);\\
\nonumber s^2 &\equiv \sum_{n = 0}^\infty n^2p_n - m^2 
= 2 p_0-(m-2)(m-1) \\ 
\label{eq:ss2}&+\sum_{n=2}^{\infty}p_n(n-2)(n-1).
\end{align}

Introducing the notation \(c \equiv \sum_{n = 2}^{\infty}p_n(n-2)\) and suppose it is a fixed value along with the mean \(m\), we obtain the equation 
\begin{equation}
\label{eq:conversion} p_1 = 2 - m + c - 2p_0
\end{equation} 
for the line representing possible states in the \(p_1 - p_0\) graph \ernew{(see Fig.~\ref{fig:NGconversion}, orange and blue lines)}. Quantum non-Gaussianity under these assumptions is equivalent to \(p_0\) being smaller than \(p_0^*(c)\), the abscissa of the intersection point of this line and the black curve prescribed by \eqref{eq:p0p1-criterion} \ernew{(which defines the section of the line within the grey, QNG area)}. It is straightforward to show that \(c' < c\) \(\Leftrightarrow\) \(p_0^*(c') < p_0^*(c)\). Therefore, \ernew{if \(p_0\) is smaller than \(p_0^*(0)\), then \(p_0<p_0^*(0) \leqslant p_0^*(c)\)} implies non-Gaussianity for all values of \(c\). \ernew{That is, we established that assuming \(c=0 \Leftrightarrow p_{3+} = 0\) is the worst-case scenario if we want to show quantum non-Gaussianity for a fixed value of the mean photon number.}


\ernew{In order to obtain a bound for the moments instead of probabilities, we} can also express \(p_0\) from Eq.~\eqref{eq:ss2}. We get that \ernew{the non-Gaussianity witness} \(p_0 < p_0^*(0)\) is equivalent to \(s^2 < 2 p_0^*(0)-(m-2)(m-1) +\sum_{n = 2}^\infty p_n(n-2)(n-1)\). Since \(\sum_{n = 2}^\infty p_n(n-2)(n-1)\geqslant 0\) (with equality for \(p_{3+} = 0\)), the state is certainly QNG if 
\begin{equation}\label{s2_prob}
s^2 < 2 p_0^*(0)-(m-2)(m-1).
\end{equation}
As a reminder, we calculate \(p_0^*(0)\) by taking the intersection point for \(c = 0\) \ernew{ (i.e., \(p_{3+} = 0\)), and the right-hand side of \eqref{s2_prob} corresponds to exactly same case.} 

In summary, we can convert the Fock probability-based witness to photon number mean and variance by assuming \(p_{3+} = 0\), which yields
\begin{align}
\label{eq:conv-m} \tilde m (r) &= \tilde p_1(r) + 2\cdot(1 - \tilde p_1(r) - \tilde p_0(r))\\
\label{eq:conv-s2} \tilde s^2(r) &= \tilde p_1(r) + 4\cdot(1 - \tilde p_1(r) - \tilde p_0(r)) - \tilde m^2(r).
\end{align}

Importantly, the conversion is only exact if \(p_{3+} = 0\), for \(p_{3+} >0\), the transformed witness \eqref{eq:conv-m}-\eqref{eq:conv-s2} is stricter than the original, \eqref{eq:p0p1-criterion}. Furthermore, the transformed witness' applicability is restricted to \(m \leqslant 2\). However, note that the original witness is also inconclusive if \(p_0 = p_1 = 0\), which is not restrictive for the moment-based witness.

\end{document}